\documentclass[sigconf,nonacm]{acmart}

\usepackage[utf8]{inputenc}
\usepackage{siunitx}
\usepackage{algorithmic}
\usepackage{textcomp}
\usepackage{xcolor}
\usepackage{url}
\usepackage{listings}
\usepackage{amsfonts}
\usepackage{amsmath}
\DeclareSymbolFontAlphabet{\mathbb}{AMSb}

\newcommand\tab[1][0.2cm]{\hspace*{#1}}

\usepackage{array, makecell} 

\usepackage[ruled,linesnumbered]{algorithm2e}
\usepackage{dblfloatfix}
\usepackage{graphicx}
\usepackage{enumitem}

\AtBeginDocument{%
  \providecommand\BibTeX{{%
    \normalfont B\kern-0.5em{\scshape i\kern-0.25em b}\kern-0.8em\TeX}}}

\begin{document}

\title{Efficient FPGA-based ECDSA Verification Engine for Permissioned Blockchains}

\author{Rashmi Agrawal}
\authornote{Work done during internship at Xilinx.}
\email{rashmi23@bu.edu}
\affiliation{%
  \institution{Integrated Circuit and Systems Group}
  \country{Boston University, USA}
}

\author{Ji Yang}
\email{jamiey@xilinx.com}
\affiliation{%
  \institution{Xilinx Research Labs}
  \country{Xilinx, USA}
}

\author{Haris Javaid}
\email{harisj@xilinx.com}
\affiliation{%
  \institution{Xilinx Research Labs}
  \country{Xilinx, Singapore}
}


\begin{abstract}
As enterprises embrace blockchain technology, many real-world applications have been developed and deployed using permissioned blockchain platforms (access to network is controlled by allowing only nodes with known identities). Such blockchain platforms heavily depend on cryptography to provide a layer of trust within the network, thus verification of cryptographic signatures often becomes the bottleneck. The Elliptic Curve Digital Signature Algorithm (ECDSA) is the most commonly used cryptographic scheme in permissioned blockchains. In this paper, we propose an efficient implementation of ECDSA signature verification on an FPGA, in order to improve the performance of permissioned blockchains that aim to use FPGA-based hardware accelerators.

In particular, we propose several optimizations for modular arithmetic (e.g., custom multipliers and fast modular reduction) and point arithmetic (e.g., significantly reduced number of point double and addition operations, and optimal width NAF representation). Based on these optimized modular and point arithmetic modules, we propose an ECDSA verification engine that can be used by any application for fast verification of ECDSA signatures. We further optimize our ECDSA verification engine for Hyperledger Fabric (one of the most widely used permissioned blockchain platforms) by moving carefully selected operations to a precomputation block, thus simplifying the critical path of ECDSA signature verification. From our implementation on Xilinx Alveo U250 accelerator board with target frequency of 250MHz, our ECDSA verification engine can perform a single verification in $760\mu s$ resulting in a throughput of $1,315$ verifications per second, which is \textasciitilde$2.5 \times$ faster than state-of-the-art FPGA-based implementations \cite{glas2011prime,knevzevic2016low}. Our Hyperledger Fabric-specific ECDSA engine can perform a single verification in $368\mu s$ with a throughput of $2,717$ verifications per second.
\end{abstract}




\maketitle


\section{Introduction}
\label{sec:introduction}
Beyond the hype, blockchain technology is emerging as one of the most disruptive technologies, with real-world use cases in many domains from digital identity management to financial services, supply chains, and product provenance. The blockchain technology essentially provides a mechanism to execute and record transactions (representative of business logic) in an immutable ledger, by grouping transactions into blocks and creating a hash-linked chain of those blocks. The nodes in a blockchain network agree upon a total order of the blocks and transactions in each block (consensus), and each node maintains its own copy of the ledger, resulting in a distributed ledger. The beauty of blockchain technology is that it seamlessly combines consensus mechanisms with cryptography to provide a layer of trust for executing and recording transactions within a network of mutually untrusting nodes.

Blockchains are generally categorized into two types. In public blockchains, such as Bitcoin and Ethereum, any node can participate in the network without a specific identity and proof-of-work based consensus is used. Proof-of-work consensus is computationally very intensive because of the massive amounts of hashes that need to be computed, and thus becomes the bottleneck. Consequently, public blockchains use hardware acceleration for hashing. For example, Bitcoin network is dominated by ASIC based nodes while GPU based nodes dominate the Ethereum network~\cite{Ridely2021Ethereum}. In permissioned blockchains, on the other hand, only nodes with known identities are part of and allowed to interact with the network, while the consensus is delegated to only a few nodes (based on BFT or CFT protocols~\cite{Bano2017}). Consequently, nodes are authenticated and transactions are validated cryptographically, thus cryptographic operations become the bottleneck rather than the consensus mechanism~\cite{Thakkar2021}. Typically, permissioned blockchains are deployed on multi-core servers to benefit from some parallelism available across processing of multiple transactions.

Since permissioned blockchains provide trust through cryptographic authentication, and data integrity and replication through distributed ledger, they are becoming increasingly popular for implementation of enterprise applications. Many permissioned blockchain platforms such as Hyperledger Fabric~\cite{HyperledgerFabric}, Quorum~\cite{Quorum} and Corda~\cite{Corda} are now available. Fabric is an open-source and enterprise-grade implementation of a permissioned blockchain, and is one of the most widely used platforms with many real-world applications already developed and deployed from finance and supply chain domains~\cite{Castillo2021,Ehrlich2021}.

\begin{figure*}[!b]
	\begin{center}
		\includegraphics[width=6.5in]{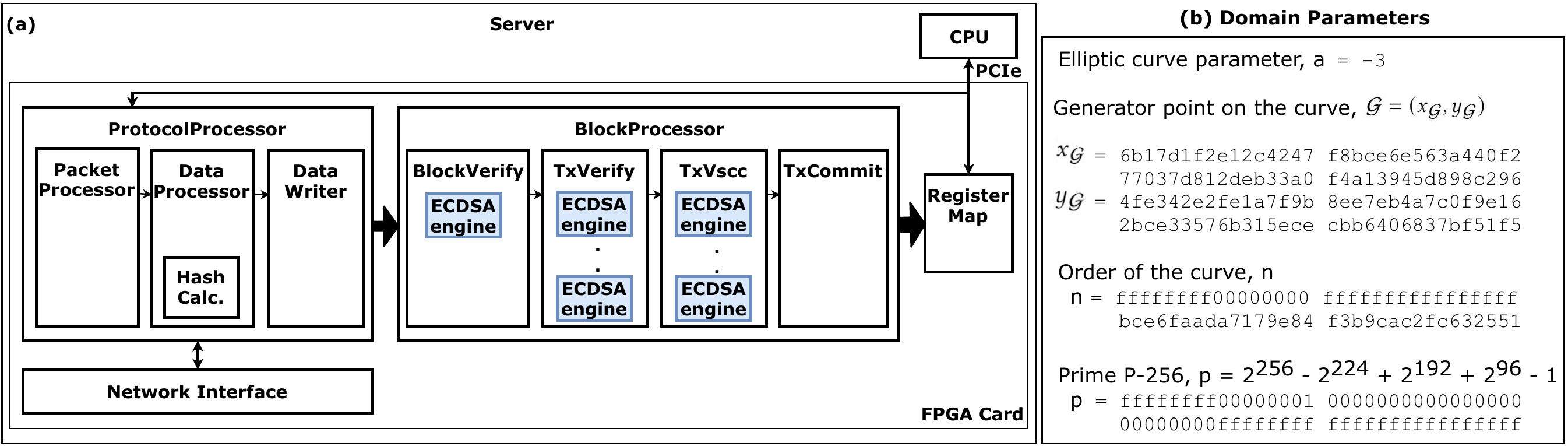} 
	\end{center}
	\caption{Blockchain Machine: (a) FPGA-based hardware accelerator, (b) Domain parameters for Hyperledger Fabric.}
	\label{fig:BCM}
\end{figure*}

In a Hyperledger (HL) Fabric network, one of the nodes is a validator peer, which is responsible for validating a block and all of its transactions, before committing that block to the ledger. Many recent works~\cite{Gorenflo2019FastFabric,Javaid2019,Chung2019,Thakkar2021} have shown that validator peer is the major bottleneck and critically affects the peak throughput. Some of these works~\cite{Javaid2019,Chung2019,Thakkar2018} further demonstrated that verification of cryptographic signatures is the major bottleneck inside a validator peer. By default, Fabric uses $256$-bit ECDSA scheme for signature generation and verification. Validation of a block involves verification of its creator's ECDSA signature, and validation of each transaction in a block involves verification of multiple ECDSA signatures (from creator and different peers in the network). Similar validation nodes exist in other permissioned blockchains such as Quorum~\cite{Quorum} and Hyperledger Besu~\cite{HyperledgerBesu} (permissioned variants of Ethereum).

Hardware acceleration was recently proposed for validation of blocks in permissioned blockchains, specifically HL Fabric. The work in~\cite{javaid2021blockchain} proposed a CPU-FPGA based system where a multi-core server with a network-attached FPGA card (connected to the CPU via PCIe) is used to accelerate validator peer of a Fabric network. All the compute-intensive operations of validation were offloaded to the FPGA accelerator, including verification of ECDSA signatures. Although the work in~\cite{javaid2021blockchain} demonstrated an order of magnitude speedup in block validation compared to CPU-only implementation, interestingly enough, the ECDSA signature verification still turned out to be the critical path in the FPGA accelerator. Therefore, in this paper, we focus on an efficient FPGA-based implementation of ECDSA signature verification, in order to improve performance of permissioned blockchains that aim to use FPGA-based accelerators.

More specifically, we focus on accelerating ECDSA signature verification over NIST $P$-$256$ elliptic curve. Working with NIST $P$-$256$ curve requires performing $256$-bit modular and point arithmetic operations. This is challenging to implement on FPGAs because $256$-bit wide multipliers, adders/subtractors, and dividers are not readily available. Therefore, a naive implementation will lead to a resource intensive design, leaving less resources for other operations of an accelerator. This, in turn, makes it challenging to fit the entire accelerator on an FPGA and meet the required timing constraints.

\noindent In particular, we make the following contributions: \vspace{-0.09in}
\begin{itemize}[leftmargin=2.5ex]
	\item \textbf{FPGA-specific optimizations for modular arithmetic:} We present algorithmic optimizations for modular arithmetic modules to enable optimized FPGA-based implementations. We specifically propose a custom $256$-bit multiplier for integer multiplication (used in modular multiplication) and a $258$-bit multiplier for Barrett reduction module. These multipliers efficiently leverage internal multipliers and registers of the DSP blocks in FPGA for better performance. We also propose an efficient algorithm to perform fast modular reduction over $P$-$256$ without using expensive $256$-bit comparators.
	
	\item \textbf{Algorithmic optimizations for point arithmetic:} We present optimizations for simultaneous-point and fixed-point multiplication algorithms (used in ECDSA verification) by reducing the overall number of point double and addition operations respectively. We use projective Chudnovsky coordinate system along with optimal width for non-adjacent form (NAF) representation to further reduce the total number of point arithmetic operations.     

	\item \textbf{HL Fabric-specific ECDSA verification:} We present a fast ECDSA signature verification engine by leveraging the fact that the generator point $\mathcal{G}$ is fixed and the public key $\mathcal{K}$ can be extracted in advance. This allows us to move a major chunk of point arithmetic from ECDSA verification to a precompute block. Consequently, the point arithmetic during actual ECDSA computation reduces to just point addition operations, resulting in a much faster signature verification.   
\end{itemize}

We implemented our optimized ECDSA verification engines on Xilinx Alveo U250 board \cite{Xilinx2020Alveo} with a target frequency of 250MHz. For modular arithmetic modules, we observe on average $1.5 \times$ speed up compared to \cite{guneysu2008ultra}, while for point arithmetic modules, we observe on average $3.2\times$ speedup compared to the state-of-the-art implementations~\cite{guneysu2008ultra,vliegen2010compact,kudithi2019high,mentens2007secure,mcivor2004fpga,schinianakis2008rns}. Our ECDSA verification engine, using these modules, performs a signature verification in $760\mu s$ resulting in a throughput of $1,315$ verifications per second, which is \textasciitilde$2.5\times$ faster than the existing FPGA-based  implementations~\cite{glas2011prime,knevzevic2016low}. With HL Fabric-specific optimizations, our ECDSA verification engine can perform a signature verification in $368\mu s$ resulting in a throughput of $2,717$ verifications per second.

\section{Background and Preliminaries}
\label{sec:background}

\subsection{Blockchain Machine}
Figure~\ref{fig:BCM}(a) depicts a simplified overview of hardware accelerator for HL Fabric that was proposed in~\cite{javaid2021blockchain}. The blocks are received in the FPGA accelerator card through the integrated network interface. The first module, ProtocolProcessor, processes the incoming packets and extracts relevant data, such as block id, number of transactions in the block, ECDSA signatures, public keys from identity certificates, etc. The second module, BlockProcessor, uses this data to validate the block and its transactions, before committing the transactions. Once a block is validated, the Fabric software running on CPU accesses validation results from hardware and continues on with committing the block to the ledger.

Internally, the BlockProcessor uses a configurable number of ECDSA verification engines distributed across multiple stages to process the ECDSA verifications as fast as possible. Each ECDSA engine accepts a verification request in the form of \{signature, key, hash\}, where signature includes both its r and s components, key is the public key of the signer with both x and y components, and hash corresponds to the hash of message (e.g., block or transaction). Typically, the hash is computed as part of the ECDSA verification, but in Blockchain Machine, it is precomputed by the ProtocolProcessor for better performance. The work in~\cite{javaid2021blockchain} reported that a single ECDSA verification was the slowest operation in their hardware accelerator. Therefore, in this paper, we design an efficient FPGA-based ECDSA verification engine that can be used inside the Blockchain Machine, or similar hardware accelerators for permissioned blockchains. 

\subsection{ECC}
\label{ECC}
\vspace{-0.05in}
An elliptic curve $\mathcal{E}$ over a prime field $\mathbb{F}_p$ is defined by a pair of tuple $(x,y)$ satisfying the Weierstrass equation 
\vspace{-0.05in}
\begin{equation}
    y^2 = x^3 + ax + b
\end{equation}

 where $a$ and $b$ belong to a Galois field $GF(p)$ with $p > 3$ and $\mathbb{F}_p = GF(p)$. Point arithmetic allows us to compute any point $\mathcal{P}_i = (x_i,y_i)$ on the elliptic curve. Addition of two points $\mathcal{P}$ and $\mathcal{Q}$, where $\mathcal{P} \ne \mathcal{Q}$, is defined by $\mathcal{R} = \mathcal{P} + \mathcal{Q}$. However, when $\mathcal{P} = \mathcal{Q}$, point addition is performed as a point double operation, resulting in $2\mathcal{P}$. Although point addition and double operations can be performed in affine or projective coordinates, affine coordinates are not preferred as they require expensive modulo inverse computations~\cite{knevzevic2016low}. In this work, we focus on performing point addition and point double operations in projective Chudnovsky coordinates~\cite{cohen1998efficient} where each point $\mathcal{P}$ is represented as a quintuple $(X,Y,Z,Z^2,Z^3)$, which corresponds to the affine point $(x = X/Z^2,y = Y/Z^3)$. These coordinates give a speed benefit over affine coordinates when the cost for modulo inversion is significantly higher than the modular multiplication. Therefore, we use projective Chudnovsky coordinates because FPGAs have DSP blocks with multipliers that can perform fast multiplications. The point addition and double equations in projective Chudnovsky coordinates are given in Table~\ref{table:PCC}. For NIST Prime curves, which include $P$-$256$ that is used in HL Fabric, domain parameters are given in FIPS $186-4$~\cite{NIST-FIPS} and are listed along with Figure~\ref{fig:BCM}(b) for reference. 

\begin{table}[h!]
\caption{Point addition and double equations in projective Chudnovsky coordinates.}
\vspace{-0.15in}
\begin{center}
  \begin{tabular}{| c | c |} 
    \hline
    \makecell{Point Addition \\ $(X_1,Y_1,Z_1,Z_1^2,Z_1^3,X_2,Y_2,Z_2,Z_2^2,Z_2^3)$} & \makecell{Point Double \\ $(X,Y,Z,Z^2,Z^3)$} \\    
    \hline
    \hline
    $U_1 = X_1 * Z_2^2$, $U_2 = X_2 * Z_1^2$ & $S = 4*X*Y^2$ \\
    $S_1 = Y_1 * Z_2^3$, $S_2 = Y_2 * Z_1^3$ & $M = 3*X^2 + a*(Z^2)^2$ \\
    $H = U_2 - U_1$, $R = S_2 - S_1$ & $X_1 = M^2 - 2*S$ \\
    $X_3 = R^2 - H^3 - 2*U_1*H^2$ & $Y_1 = M*(S - X_1) - 8*Y^4$ \\
    $Y_3 = R*(U_1*H^2 - X_3) - S_1*H^3$ & $Z_1 = 2*Y*Z$ \\
    $Z_3 = H*Z_1*Z_2$ & $Z_1^2 = Z_1^2$ \\
    $Z_3^2 = Z_3^2$, $Z_3^3 = Z_3^2*Z_3$ & $Z_1^3 = Z_1^2 * Z_1$ \\
    Return$(X_3,Y_3,Z_3,Z_3^2,Z_3^3)$ & Return$(X_1,Y_1,Z_1,Z_1^2,Z_1^3)$ \\
    \hline
\end{tabular}
\label{table:PCC}
\vspace{-0.15in}
\end{center}
\end{table}

\section{Related Work}
\label{sec:related-works}
In this section, we discuss the most relevant exisiting works in the literature. There are many publications focusing on acceleration of modular multiplication, point arithmetic for ECC, or ECDSA signature verification in hardware. Overall, these hardware accelerators can be broadly classified into three categories based on their implementation platforms. First category includes reconfigurable architectures that are FPGA-based~\cite{guneysu2008ultra,vliegen2010compact,glas2011prime,sghaier2017design,tachibana2019fpga,kudithi2019high,sau2021binary}, the second category includes $8$-bit AVR-based implementations~\cite{gura2004comparing,liu2015efficient,park2020fast} for embedded devices that are resource constrained, and the third category includes ASIC-based hardwired architectures~\cite{youssef2014low,knevzevic2016low,ji2018asic} to achieve high-performance for a specific elliptic curve. 

Some of the works~\cite{sau2021binary,youssef2014low} focus on elliptic curves over binary fields $GF(2^m)$ only. This is because the hardware implementation of binary field operations results in a carry-free logic, and thus these fields are the most optimal for use in hardware in terms of both speed and area. Other works focus on elliptic curves over prime fields $GF(p)$, but perform hardware acceleration for a different field size like $163$-bits~\cite{sghaier2017design}, $192$-bits~\cite{liu2015efficient}, or $224$-bits~\cite{gura2004comparing}, which is not of interest to blockchain platforms that typically use $256$-bits prime field. There is also a wide range of work~\cite{guneysu2008ultra,vliegen2010compact,kudithi2019high,park2020fast} that focuses on accelerating only point arithmetic for ECC over NIST $P$-$256$ elliptic curve, and do not implement the entire ECDSA verification. 

Tachibana et al.~\cite{tachibana2019fpga} accelerate ECDSA verification algorithm over Secp256k1 elliptic curve for Bitcoin on an Intel Cyclone IV FPGA. Their single ECDSA verification takes about $145.52$~ms. Glas et al.~\cite{glas2011prime} present an FPGA-based ECDSA core for $256$-bits field using Xilinx Virtex $5$ FPGA board. The authors integrate their hardware core in a vehicle-to-vehicle communication system and compare the performance against microcontroller-based implementation. Ji et al.~\cite{ji2018asic} and Kne{\v{z}}evi{\'c} et al.~\cite{knevzevic2016low} implement the ECDSA verification algorithm over NIST $P$-$256$ elliptic curve as ASICs. However, Ji et al.~\cite{ji2018asic} did not report any latency numbers. We will present the results from both~\cite{glas2011prime} and~\cite{knevzevic2016low} for comparison in Section~\ref{sec:evaluation}.

In this work, we optimize modular arithmetic operations (specifically multiplications) to utilize DSP blocks on FPGA and accelerate point arithmetic operations over NIST $P$-$256$ elliptic curve, and then combine these modules to implement an efficient ECDSA verification engine (both generic and HL Fabric-specific). To the best of our knowledge, our proposed architecture performs the fastest ECDSA verification over NIST $P$-$256$ prime field on an FPGA (see Section~\ref{sec:evaluation}).

\vspace{-2ex}
\section{ECDSA Verification Algorithm}
\label{sec:ECDSAverf}

A sender sends the message digest $z = H(m)$, the signature $(r,s)$, and the public key $\mathcal{K} = (x_{\mathcal{K}},y_{\mathcal{K}})$, which are verified by the receiver. It is assumed that the receiver knows the ECC domain parameters ($a$, $\mathcal{G} = (x_{\mathcal{G}}, y_{\mathcal{G}})$, $n$, and $p$ described in Section~\ref{ECC}). Algorithm~\ref{alg:ECDSA} defines the ECDSA signature verification process, where the output indicates whether the signature is valid or not. Typically, an ECDSA verification algorithm computes the message digest, however we assume that the hash is precomputed to be aligned with the Blockchain Machine architecture (the hash is computed by the ProtocolProcessor module). Hence, the algorithm receives the computed hash instead of the message. It is also worth noting that the signature verification algorithm performs modular reductions with respect to two different primes. For point arithmetic (line $7$), the prime $p$ is used, whereas for the rest of the operations (lines $4$-$6$ and $8$) order of the curve $n$ is used.    

\begin{figure}[t]
	\begin{center}
		\includegraphics[width=1.00\columnwidth]{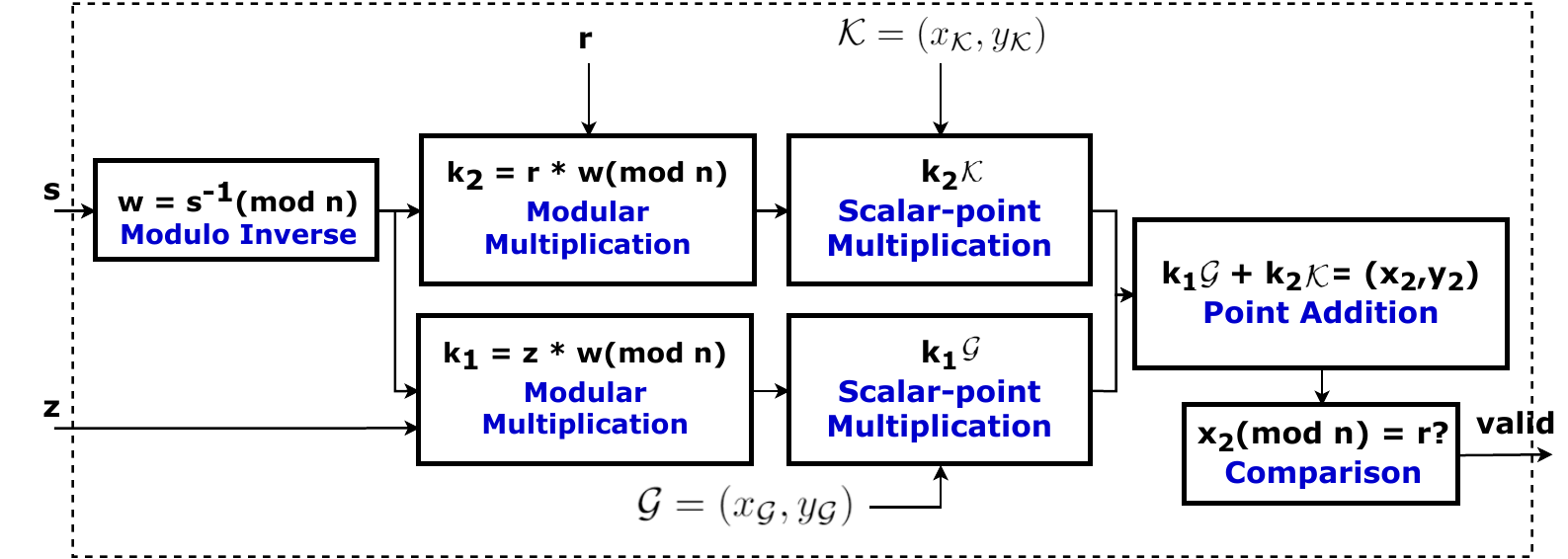}
	\end{center}
	\vspace{-0.15in}
	\caption{Operations in ECDSA signature verification.}
	\vspace{-0.10in}
	\label{fig:ECDAoperations}
\end{figure}

When we translate Algorithm~\ref{alg:ECDSA} into a pictorial representation as shown in Figure~\ref{fig:ECDAoperations}, it is evident that we need to implement modular arithmetic operations such as modulo inverse, multiplication and reduction followed by various point arithmetic operations such as point addition, double and scalar-point multiplication.

\begin{algorithm}[!t]
	\caption{ECDSA Verification}
	\begin{algorithmic}[1]
		\renewcommand{\algorithmicrequire}{\textbf{Input:}}
		\renewcommand{\algorithmicensure}{\textbf{Output:}}
		\REQUIRE Message digest $z$, the signature $(r,s)$, \\ and the public key $\mathcal{K} = (x_{\mathcal{K}},y_{\mathcal{K}})$
		\ENSURE  Valid or invalid
		\IF {$(r,s)$ not in range $[1,n-1]$}
		\STATE Return(Invalid)
		\ENDIF
		\STATE Compute $w = s^{-1} \mod n$
		\STATE Compute $k_1 = z*w \mod n$
		\STATE Compute $k_2 = r*w \mod n$
		\STATE Compute $(x_2,y_2) = k_1G + k_2\mathcal{K}$
		\IF {$r == x_2 \mod n$}
		\STATE Return(Valid)
		\ELSE
		\STATE Return(Invalid)
		\ENDIF
	\end{algorithmic}
	\label{alg:ECDSA}
\end{algorithm}

\section{FPGA-based ECDSA Verification Engine}
\label{sec:architecture}
In this section, we present the architecture of our efficient ECDSA verification engine. We would like to highlight here that the ECDSA signature verification algorithm deals with the information that is publicly known, therefore there is no secret information to leak through side-channels while performing a signature verification. This widens our choice of algorithms for implementing various modules within the ECDSA verification engine. Therefore, in our hardware implementation, we select algorithms that utilize minimal hardware resources while resulting in low latency.

\subsection{Modular Arithmetic}
\label{FA}
To efficiently implement $256$-bit wide modular arithmetic on an FPGA, we implement all modular arithmetic modules using multi-word integer arithmetic~\cite{HG2006}. In multi-word arithmetic, a $256$-bit field element $a$ can be represented as 
\vspace{-0.05in}
\begin{equation}
    a = 2^{(t-1)W}A[t-1] + \dots + 2^{2W}A[2] + 2^WA[1] + A[0]
\end{equation}
where, $W$ and $t=256/W$ define the word length and the number of words to operate on respectively. We set the value of $W$ diligently for every operation to efficiently utilize the underlying resources of an FPGA. For example, on Xilinx Alveo U250 FPGA board, the multipliers are $16$-bit wide and hence, $W$ can be set as $16$. The advantages of using multi-word integer approach and setting the value of $W$ carefully are two-fold. First, it helps in performing the modular operations using hardwired DSP blocks on the FPGA. Second, it helps in achieving a higher maximum operating frequency for the entire design. \\

\noindent\textbf{Modular Subtraction:}
Subtraction in $\mathbb{F}_{p}$ can be performed using Algorithm~\ref{alg:MS}~\cite{HG2006} with multi-word integer approach. The adder/subtractor in DSP blocks have $48$-bit wide inputs. However, for multi-word integer approach, the maximum bit width we can use is $32$~bits ($32$ being the largest integer dividing $256$ symmetrically).  Therefore, to use DSP for subtraction, we set the parameter $W$ as $32$ and thus, we have $t=8$ words to operate on. Along with each subtraction operation, we subtract the previous carry bit and store the next carry bit (line $3$). As we operate on a single word at a time, we utilize only one DSP block to implement the modular subtraction operation. We want to highlight here that point arithmetic (including point double and add) in projective Chudnovsky coordinate requires only one modular addition operation (see Table~\ref{table:PCC}). Therefore, we decided to perform the modular addition using our modular subtraction module with $2'$s complement input for the second operand. Implementing a separate adder leads to inefficient resource utilization as the adder will remain idle for most of the time. \\

\vspace{-0.15in}
\begin{algorithm}[ht]
 \caption{Subtraction in $\mathbb{F}_p$}
 \begin{algorithmic}[1]
 \renewcommand{\algorithmicrequire}{\textbf{Input:}}
 \renewcommand{\algorithmicensure}{\textbf{Output:}}
 \REQUIRE Modulus $p$, integers $a,b \in [0,p-1]$, $t=8$
 \ENSURE  $c = (a-b) \mod p$
  \STATE Set (A[t-1],\dots,A[0]) $\leftarrow$ $a$, (B[t-1],\dots,B[0]) $\leftarrow$ $b$
  \STATE ($carry$, $C[0]$) $\leftarrow$ $A[0] - B[0]$
  \FOR{$i$ from $1$ to $t-1$}
    \STATE ($carry$, $C[i]$) $\leftarrow$ $A[i] - B[i] - carry$
  \ENDFOR
  \IF {$carry$ = $1$}
    \STATE add $p$ to $c = (C[t-1],\dots,C[2],C[1],C[0])$ 
  \ENDIF
  \STATE Return($c$)
 \end{algorithmic}
 \label{alg:MS}
 \end{algorithm}

\noindent\textbf{Integer Multiplication Module:}
We adopt a hybrid approach to integer multiplication through a combined schoolbook~\cite{rafferty2017evaluation} and Karatsuba-Ofman~\cite{HG2006} approach. The motivation behind adopting a hybrid approach is as follows. 
Our target FPGA board has DSP blocks with $27$x$18$ bit wide multipliers. However, both multiword arithmetic and schoolbook algorithm require operands to be split symmetrically (i.e., both operands must have the same base). Therefore, the maximum we can use is $16$x$16$ bit wide multiplier to multiply two $256$-bit operands ($16$ being the largest number that can split $256$ symmetrically). If we set $W$ as $16$, we will have $t=16$ words to operate on, thus implementing just the schoolbook multiplication algorithm~\cite{HG2006} will lead to a high latency (at least $256$~clock cycles), which is not acceptable as many multiplications are performed in point arithmetic.
Therefore, the use of hybrid approach lowers the latency of a multiplication performed using multi-word approach with schoolbook multiplication. The steps of our proposed approach are shown in Algorithm~\ref{alg:Mult}.

\begin{figure}[ht]
  \begin{center}
  \vspace{-0.15in}
    \includegraphics[width=1\columnwidth]{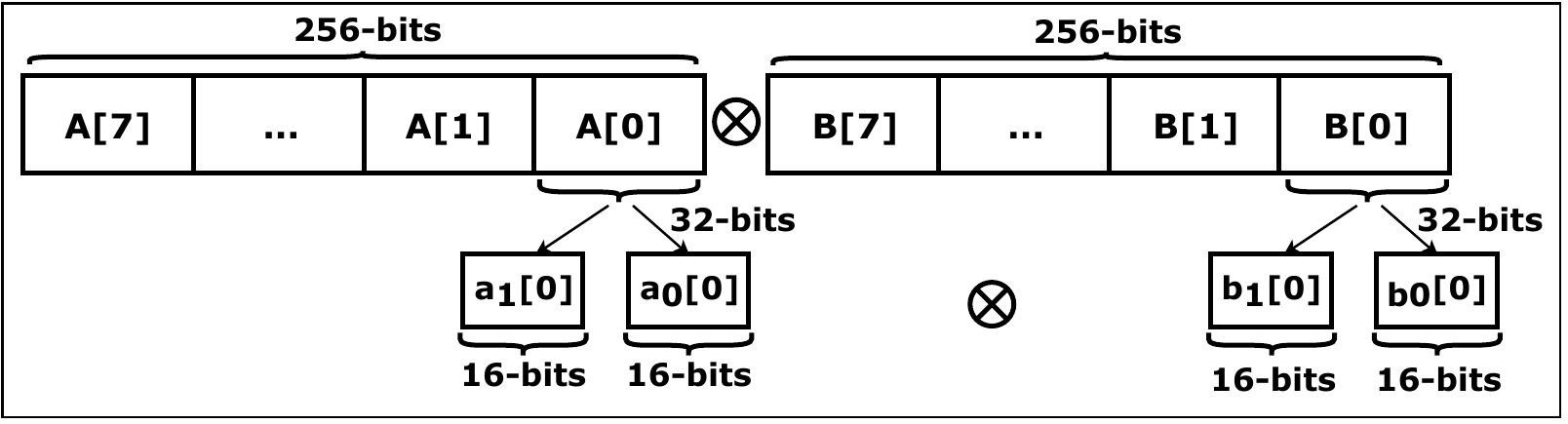}
  \end{center}
  \vspace{-0.15in}
  \caption{Schoolbook+Karatsuba-Ofman multiplication approach.}
  \vspace{-0.10in}
  \label{fig:IntMult}
\end{figure}

We first take the schoolbook multiplication algorithm and set the parameter $W$ as $32$ (twice the input width of a multiplier in DSP block) to split $256$-bit operands into $t=8$ $32$-bit words. Next, we target the $32$-bit operands and split them into $16$-bit operands (see Figure~\ref{fig:IntMult}) and multiply these $16$-bit operands using Karatsuba-Ofman equation as follows:
\begin{equation}
    ab = (a_1b_12^{2l} + [(a_0+a_1)(b_0+b_1)-a_1b_1-a_0b_0]2^l + a_0b_0)
\end{equation}

where $l$ is $16$. Moreover, as multiplication operations (line $7$ in Algorithm~\ref{alg:Mult}) are independent, we unroll the loops in the algorithm to perform multiplications and accumulations in parallel. As a result, we can perform $256$-bit integer multiplications in just $39$ clock cycles using the schoolbook multiplication algorithm with the Karatsuba-Ofman split
to utilize the available wide multipliers in DSP blocks efficiently.
Note that various existing works~\cite{zhou2009improving,zutter2009acceleration,xie2017efficient} (implementing different cryptographic schemes) exploit Karatsuba-ofman/Karatsuba algorithm to perform modular multiplication on an FPGA. Therefore, novelty of this work lies in the way we combine schoolbook multiplication and Karatsuba-ofman algorithm together to implement a low-latency, parallel multiplier. 
\\

\vspace{-0.05in}
\begin{algorithm}[t]
	\caption{Hybrid Integer Multiplication}
	\begin{algorithmic}[1]
		\renewcommand{\algorithmicrequire}{\textbf{Input:}}
		\renewcommand{\algorithmicensure}{\textbf{Output:}}
		\REQUIRE Integers $a,b \in [0,p-1]$, $t=8$, $l=16$
		\ENSURE  $c = (a.b)$
		\STATE Set (A[t-1],\dots,A[0]) $\leftarrow$ $a$, (B[t-1],\dots,B[0]) $\leftarrow$ $b$
		\STATE Set C[i] $\leftarrow$ $0$ for $0 \le i \le 2t-1$
		\FOR{i from $0$ to $t-1$}
		\STATE U $\leftarrow$ 0
		\FOR{j from $0$ to $t-1$}
		\STATE $(a_1,a_0)$ $\leftarrow$ A[i], $(b_1,b_0)$ $\leftarrow$ B[i]
		\STATE $ab = a_1b_12^{2l} + [(a_0+a_1)(b_0+b_1)-a_1b_1-a_0b_0]2^l + a_0b_0$
		\STATE (U,V) $\leftarrow$ C[i+j] + ab + U
		\STATE C[i+j] $\leftarrow$ V
		\ENDFOR
		\STATE C[i+t] $\leftarrow$ U 
		\ENDFOR
		\STATE Return(c)
	\end{algorithmic}
	\label{alg:Mult}
\end{algorithm}

\noindent\textbf{Fast $P$-$256$ Modular Reduction Module:} The result of an integer multiplication yields a $512$ bits result that needs to be reduced to $256$ bits by performing modular reduction. In point arithmetic, modular reduction is performed using the prime $p=$ $P$-$256$ while other modular reductions in ECDSA verification engine are performed using the other prime, i.e., the order of the curve $n$. Modular reduction with $n$ follows this discussion. Either ways, $256$-bit modular reduction is very expensive to implement in hardware due to large division operation. We observe that a single modular reduction using the $\%$~operator in Verilog requires about $57,000$~LUTs on the target FPGA, which is quite expensive. Therefore, efficient implementation of modular reduction is crucial to the design of an ECDSA verification engine.

NIST recommends a fast modulo reduction $P$-$256$ algorithm~\cite{HG2006} as $p=$ $P$-$256$ is a general Mersenne prime~\cite{solinas1999generalized}. The algorithm replaces large division operations with simple additions and subtractions by exploiting the structure of this Mersenne prime. Therefore, by using this algorithm, we can perform the $256$-bit modular reduction using two left shifts (multiplication by $2$), four additions, and four subtractions. However, the result generated from this algorithm can be in the range $-4p$ to $5p$ instead of $0$ to $p$. So we need to perform a correction by either adding to the result or subtracting from the result, a suitable value of $p$ within this range. This correction step, however, requires performing many $256$-bit comparisons to figure out the exact range in which the result lies. On FPGAs, a $256$-bit comparator leads to long carry chains impacting the timing constraints of the design adversely. Therefore, to avoid performing many large parallel comparisons at once, we check the results immediately after each step of the computation. The steps of our proposed fast modulo reduction $P$-$256$ algorithm are shown in Algorithm~\ref{alg:MR}. We can see that on line $2$ of the algorithm as we perform addition, we perform an immediate correction by comparing the result of addition with $p$. Similarly, after subtraction operation (lines $10$, $12$, $14$, and $16$), we compare the result with $0$ to see if it is negative and correct it accordingly.  

\begin{algorithm}[!t]
 \caption{Fast Reduction Modulo $P$-${256}$ Algorithm}
 \begin{algorithmic}[1]
 \renewcommand{\algorithmicrequire}{\textbf{Input:}}
 \renewcommand{\algorithmicensure}{\textbf{Output:}}
 \REQUIRE A $512$-bit integer $c = (c_{15},\dots,c_2,c_1,c_0)$ in base $2^{32}$ 
 \ENSURE  $r = c \mod P$-${256}$
  \STATE $s_1 = (c_7,c_6,c_5,c_4,c_3,c_2,c_1,c_0)$, $s_2 = (c_{15},c_{14},c_{13},c_{12},c_{11},0,0,0)$
  \STATE $r = s_1 + s_2$, $r = (r \ge p)$ $?$  $r - p$ : $r$
  \STATE $r = r + s_2$, $r = (r \ge p)$ $?$  $r - p$ : $r$
  \STATE $s_3 = (0,c_{15},c_{14},c_{13},c_{12},0,0,0)$
  \STATE $r = r + (s_3 << 1)$, $r = (r \ge p)$ $?$  $r - p$ : $r$
  \STATE $s_4 = (c_{15},c_{14},0,0,0,c_{10},c_{9},c_{8})$
  \STATE $r = r + s_4$, $r = (r \ge p)$ $?$  $r - p$ : $r$
  \STATE $s_5 = (c_8,c_{13},c_{15},c_{14},c_{13},c_{11},c_{10},c_{9})$
  \STATE $r = r + s_5$, $r = (r \ge p)$ $?$  $r - p$ : $r$
  \STATE $s_6 = (c_{10},c_{8},0,0,0,c_{13},c_{12},c_{11})$
  \STATE $r = r - s_6$, $r = (r < 0)$ $?$  $r + p$ : $r$
  \STATE $s_7 = (c_{11},c_{9},0,0,c_{15},c_{14},c_{13},c_{12})$
  \STATE $r = r - s_7$, $r = (r < 0)$ $?$  $r + p$ : $r$
  \STATE $s_8 = (c_{12},0,c_{10},c_{9},c_{8},c_{15},c_{14},c_{13})$
  \STATE $r = r - s_8$, $r = (r < 0)$ $?$  $r + p$ : $r$
  \STATE $s_9 = (c_{13},0,c_{11},c_{10},c_{9},0,c_{15},c_{14})$
  \STATE $r = r - s_9$, $r = (r < 0)$ $?$  $r + p$ : $r$
  \STATE Return$(r)$ 
 \end{algorithmic}
 \label{alg:MR}
 \end{algorithm}

\begin{algorithm}[!t]
 \caption{Efficient Comparison with $P$-$256$}
 \begin{algorithmic}[1]
 \renewcommand{\algorithmicrequire}{\textbf{Input:}}
 \renewcommand{\algorithmicensure}{\textbf{Output:}}
 \REQUIRE A $257$-bit integer $r$ with $0 \le r < 2p$, conditions $C_0$, $C_1$, \\ $C_2$, and $C_3$
 \ENSURE  $r \ge p$
  \STATE $r_0 = \& r[95:0]$, $r_1 = r[191:96]$, $r_2 = r[223:192]$, $r_3 = \& r[255:224]$, $r_4 = r[256]$
  \IF{($r_4 == 1$ or ($r_3 == C_3$ and $r_2 > C_2$ or ($r_2 == C_2$ and $r_1 > C_1$)))}
    \STATE Return(greater)
  \ELSIF{($r_4 == 0$ and $r_3 == C_3$ and $r_2 == C_2$ and $r_1 == C_1$ \\ and $r_0 == C_0$)}
    \STATE Return(equal)
  \ENDIF
 \end{algorithmic}
 \label{alg:COMP}
\end{algorithm}

From Algorithm~\ref{alg:MR}, it is evident that we avoid using many $256$-bit wide comparators in parallel, but we still need a $256$-bit comparator to check if the intermediate result is $\ge p$ or not. To avoid doing so, we further exploit the structure of the Mersenne prime and propose an efficient algorithm to perform this comparison without actual $256$-bit comparators. Our Algorithm~\ref{alg:COMP} is based on the following observation. We can split $P$-$256$ into four parts as follows: \\

\noindent \tab  $P_0 = P[95:0] = ffffffffffffffffffffffff$ \\
\tab  $P_1 = P[191:96] = 0$, $P_2 = P[223:192] = 1$ \\
\tab  $P_3 = P[255:224] = ffffffff$ \\
\\
\noindent Using these four parts, we can generate the following four conditions: 
\noindent \tab $C_0: \&P_0 = 1$, $C_1: P_1 = 0$, $C_2: P_2 = 1$, $C_3: \&P_3 = 1$ \\
Here, an AND ($\&$) reduction on $P_0$ gives a $1$, $P_1$ is $0$, $P_2$ is $1$, and again an $\&$ reduction on $P_3$ gives a $1$. The algorithm starts by splitting the input integer $r$ in a similar fashion as $P$-$256$ and an additional $r_4$ for the $257th$ bit (as the input integer is of $257$-bits). First, we check if $r_4$ is $1$, then $r$ is definitely greater. However, if $r_4$ is $0$, we need to check for other conditions. If $r_2 > 1$ or $r_1 > 0$ (line $3$), then we know that $r > p$ else if all the four conditions on line $5$ are satisfied, then $r = p$. This algorithm converts $256$-bit wide comparisons to four $1$-bit comparisons and can be efficiently implemented in hardware using bit-slicing and unary $\&$ operator.    \\

\noindent\textbf{Modular Reduction over $n$ using Barrett Reduction:}
We propose to use the standard Barrett reduction algorithm~\cite{HG2006} for modulo reduction over the prime $n$. Barrett reduction does not exploit the structure of $n$ but computes $r = z \mod n$, by computing a value $\hat{q}$, which when multiplied with $n$ and subtracted from $z$ will give the desired modular reduction value $r$. The algorithm requires selecting a base $b$, which when chosen as a power of two gives an efficient implementation in hardware. We select $b$ as $4$ and we precompute the parameters of Barrett reduction ($k$ and $\mu$) as these parameters are fixed and do not change at any point in computation. Our modified hardware-friendly version of the Barrett reduction is shown in Algorithm~\ref{alg:BR}. In our hardware implementation, the division operations are performed using right shift operation (line $1$) and modular reductions are performed using AND operation (line $2$). This optimization is possible because $b$ is a power-of-$2$ and for powers-of-$2$, modular reduction can be efficiently done by masking the lower-order bits using AND operation. Thus, we avoid all large division operations. However, we need to perform two large $258$-bit multiplications as the parameter $\mu$ is a $258$-bit integer. 

\begin{algorithm}[!t]
 \caption{Hardware-friendly Barrett Reduction}
 \begin{algorithmic}[1]
 \renewcommand{\algorithmicrequire}{\textbf{Input:}}
 \renewcommand{\algorithmicensure}{\textbf{Output:}}
 \REQUIRE $n$, $b = 4$, $k = \lfloor \log_b n \rfloor$ + 1, $0 \le z < b^{2k}$, and $\mu = \lfloor \frac{b^{2k}}{n} \rfloor$
 \ENSURE  $z \mod n$
  \STATE $\hat{q} \leftarrow (z >> 2(k-1)). (\mu >> 2(k+1))$
  \STATE $r \leftarrow z\ \&\ (b^{k+1}-1) - \hat{q}.n\ \&\ (b^{k+1}-1)$
  \IF{($r < 0$)}
    \STATE $r \leftarrow r + b^{k+1}$
  \ENDIF    
  \WHILE{($r \ge n$)}
    \STATE $r \leftarrow r - n$
  \ENDWHILE
  \STATE Return$(r)$
 \end{algorithmic}
 \label{alg:BR}
 \end{algorithm}

\begin{figure}[!b]
  \begin{center}
  \vspace{-0.15in}
    \includegraphics[width=1\columnwidth]{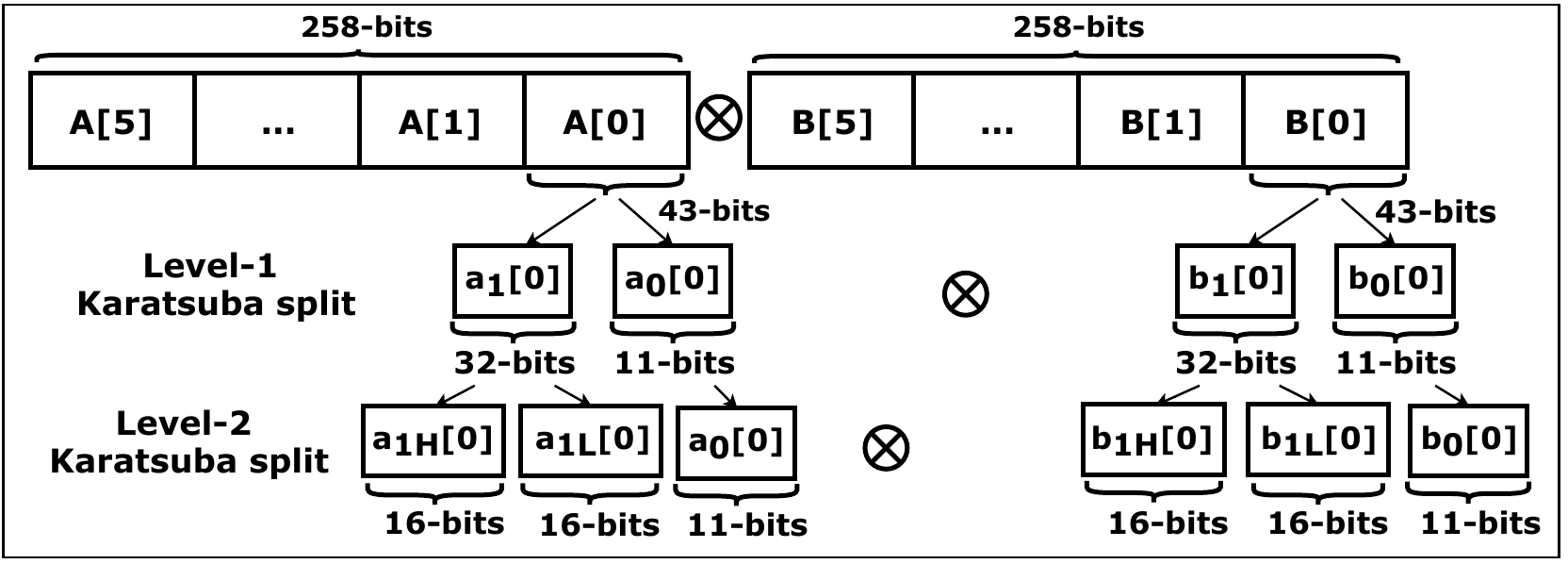}
  \end{center}
  \vspace{-0.15in}
  \caption{Proposed $258$-bit multiplication approach.}
  \vspace{-0.10in}
  \label{fig:BRMult}
\end{figure}

For $258$-bit multiplication, we again use the hybrid approach proposed in Algorithm~\ref{alg:Mult}. However, we set the parameter $W=6$ and split the input operands into $43$-bit multi-word integers. In addition, instead of a single Karatsuba-Ofman split as in Algorithm~\ref{alg:Mult}, we perform a two-level Karatsuba-Ofman split to efficiently leverage the DSP blocks for $43$-bit multiplications. At level-$1$, we split $43$-bit operands into $32$- and $11$-bit integers. Then at level-$2$, we again split $32$-bit integers into $16$-bit integers (see Figure~\ref{fig:BRMult}). Thus, effectively we perform only $16$-bit $\times\ 16$-bit, $16$-bit $\times\ 11$-bit, and $11$-bit $\times\ 11$-bit multiplications instead of $43$-bit multiplication. For example, a $43$-bit multiplication can be performed as follows: \\

\noindent \tab \textbf{Step-$1$:} Split $A[42:0]$ into $a_0 \leftarrow A[10:0]$ and $a_1 \leftarrow A[42:11]$ \\
\tab \textbf{Step-$2$:} Split $B[42:0]$ into $b_0 \leftarrow B[10:0]$ and $b_1 \leftarrow B[42:11]$ \\
\tab \textbf{Step-$3$:} $U_0,c_0 \leftarrow a_0 * b_0$ \ // $11$-bit$\times 11$-bit multiplication \\ 
\tab \textbf{Step-$4$:} $U_1,c_1 \leftarrow (a_0 * b_1) + U_0$ \ // $11$-bit$\times 32$-bit multiplication \\ 
\tab \textbf{Step-$5$:} $U_2,c_2 \leftarrow c_1 + (a_1 * b_0)$ \ // $11$-bit$\times 32$-bit multiplication \\ 
\tab \textbf{Step-$6$:} $U_3,c_3 \leftarrow (a_1 * b_1) + U_1 + U_2$ \ // $32$-bit$\times 32$-bit multiplication \\ 
\tab \textbf{Step-$7$:} $c \leftarrow (U_3,c_3,c_2,c_0)$ \ // $86$-bit multiplication result \\

\noindent In steps-$4$ and $5$ above, we need to perform an $11$-bit $\times$ $32$-bit multiplication, which we further simplify as follows: \\

\noindent \tab \textbf{Step-$1$:} Assign $a_{1lower} \leftarrow a_1[15:0]$ and $a_{1higher} \leftarrow a_1[31:16]$ \\
\tab \textbf{Step-$2$:} Pad a $0$ (in MSB) to $b_0$ to make it $12$-bits \\
\tab \textbf{Step-$3$:} $r_{lower} \leftarrow a_{1lower} * b_0$ \ // $r_{lower}$ is $28$-bits \\
\tab \textbf{Step-$4$:} $r_{higher} \leftarrow a_{1higher} * b_0$ \ // $r_{higher}$ is $28$-bits \\
\tab \textbf{Step-$5$:} $r_{mid} \leftarrow r_{lower}[27:16] + r_{higher}[11:0]$ \ // $r_{mid}$ is $13$-bits \\
\tab \textbf{Step-$6$:} $r_{higher} \leftarrow r_{higher}[26:12] + r_{mid}[12]$ \\
\tab \textbf{Step-$7$:} $result \leftarrow (r_{higher}[26:12],r_{mid}[11:0],r_{lower}[15:0])$ \\

Note that in step-$5$ above, $r_{mid}$ is $13$ bits instead of $12$ bits to account for an additional carry bit after addition. Note that although the steps listed above are specific to a $43$-bit multiplier, they can be generalized to any bit-width that is not a multiple of the DSP multiplier's input bit-width (assumed $16$ in our case) and then can be efficiently implemented leveraging the DSP blocks.\\


\noindent\textbf{Modulo Inverse Module:}
NIST recommends using Extended Euclidean algorithm~\cite{NIST-FIPS} to compute modulo inverse in ECDSA verification algorithm. However, extended Euclidean algorithm is expensive to implement in hardware requiring $256$-bit division and multiplication to compute quotient and remainder respectively. Second popular option, the Fermat’s theorem~\cite{xiang2012algebra} is comparatively efficient to implement in hardware as it requires only multiplications. However, it involves $267$~modular multiplications leading to a higher latency. Fermat's theorem approach is more suitable when a side-channel resistant implementation (computation time is independent of the input) is a must.

We choose to implement an optimized, faster modulo inverse algorithm proposed by Chen and Qin~\cite{chen2009fast}. This algorithm is suitable for hardware implementation as it has very low resource footprint and also incurs a low latency. The algorithm computes a modulo inverse using only right shift and addition operations and at any given time only two $256$-bit adders are operating in parallel. We observe that the latency of this algorithm ranges from $35$-$600$~clock cycles depending on the input, however for real test cases, the latency averages close to $550$~clock cycles across multiple evaluations. It is worth noting that we modified the actual algorithm to take modulus as an input because we leverage the same algorithm to perform modulo inverse with respect to $n$ (line 4 of Algorithm~\ref{alg:ECDSA}) as well as to convert projective Chudnovsky coordinates back to affine coordinates wherein we need to compute modulo inverse with respect to $p$.  

\subsection{Point Arithmetic}
\label{PA}
During ECDSA signature verification, we need to perform two scalar-point multiplications and one point addition operation (line 7 in Algorithm~\ref{alg:ECDSA}). We first present our generic approach to point arithmetic that can be leveraged in any application requiring fast ECDSA signature verification. We will discuss our optimized Hyperledger Fabric-specific point arithmetic approach later in  Section~\ref{sec:fabric}.  

We leverage the simultaneous-point multiplication (SPM) Algorithm~\ref{alg:SPM}, also known as “Shamir’s trick”, to operate on both the generator point ($\mathcal{P}$) and public key coordinates ($\mathcal{Q}$) at the same time. In addition, this algorithm eliminates the need to perform the point addition separately. We conducted an analysis on how the number of operations varies when different point representations such as binary, NAF, joint-sparse form (JSF), and width-$w$ NAF are used in Algorithm~\ref{alg:SPM}. Table~\ref{table:count} shows that point double operations largely remain the same while point addition operations can be reduced to as low as 112 when width-$w$ NAF is used. Note that width-$w$ NAF conversion can be done in hardware using the algorithm-$3.35$ from~\cite{HG2006}, and is trivial in comparison to point arithmetic. We use $w = 4$ to keep the storage requirements minimal.

\begin{algorithm}[t!]
 \caption{Width-$w$ NAF method for SPM}
 \begin{algorithmic}[1]
 \renewcommand{\algorithmicrequire}{\textbf{Input:}}
 \renewcommand{\algorithmicensure}{\textbf{Output:}}
 \REQUIRE Width $w$, integers $k_1$ and $k_2$, points $\mathcal{P}$ and $\mathcal{Q}$
 \ENSURE $k_1\mathcal{P} + k_2\mathcal{Q}$  
  \STATE Compute: $i\mathcal{P}$ and $i\mathcal{Q}$ for $i \in \{1,3,\dots,2^w-1\}$
  \STATE Compute $NAF_w(k_1)$ and $NAF_w(k_2)$
  \STATE $l = max\{\ell_1,\ell_2\}$ where $\ell_1$ and $\ell_2$ are lengths of $NAF_w(k_1)$ and $NAF_w(k_2)$
  \STATE $\mathcal{A} = \infty$
  \FOR{$i$ from $l-1$ down to $0$}
    \STATE $\mathcal{A} = 2\mathcal{A}$
     \IF{$k_1[i] > 0$}
      \STATE $\mathcal{A} = \mathcal{A} + k_1[i]\mathcal{P}$
     \ELSE 
      \STATE $\mathcal{A} = \mathcal{A} - k_1[i]\mathcal{P}$
     \ENDIF
     \IF{$k_2[i] > 0$}
      \STATE $\mathcal{A} = \mathcal{A} + k_2[i]\mathcal{Q}$
     \ELSE 
      \STATE $\mathcal{A} = \mathcal{A} - k_2[i]\mathcal{Q}$
     \ENDIF
  \ENDFOR
  \STATE Return$(\mathcal{A})$
 \end{algorithmic}
 \label{alg:SPM}
 \end{algorithm}

\begin{table}[t!]
\caption{Number of point arithmetic operations in SPM.}
\vspace{-0.15in}
\begin{center}
  \begin{tabular}{ | l | c | c |} 
    \hline
    Representation & Point double & Point add\\    
    \hline
    \hline
    Binary & $256$ & $193$\\
    \hline
    NAF & $256$ & $148$ \\
    \hline
    JSF & $256$ & $130$ \\
    \hline
    width-$w$ NAF ($w = 4$) & $257$ & \textasciitilde$112$ \\
    \hline
\end{tabular}
\label{table:count}
\vspace{-0.15in}
\end{center}
\end{table}

In Algorithm~\ref{alg:SPM}, point double operation (line 6) and point additions (lines 8, 10, 13, 15) cannot be done in parallel as they depend on each other. However, we reduce the number of these operations by computing various values of $\mathcal{G}$  ($3\mathcal{G},5\mathcal{G},7\mathcal{G},9\mathcal{G},11\mathcal{G},13\mathcal{G},15\mathcal{G}$) offline and storing them in BRAM because $\mathcal{G}$ is known in advance. This requires $1120$~bytes of storage space but reduces the computation of line $1$ to only $\mathcal{K}$   ($3\mathcal{K},5\mathcal{K},7\mathcal{K},9\mathcal{K},11\mathcal{K},13\mathcal{K},15\mathcal{K}$) in hardware. For these values, $3\mathcal{K}$ is computed by performing a point double on $\mathcal{K}$ followed by a point addition. We store the $2\mathcal{K}$ value temporarily and reuse it; for example, $5\mathcal{K}$ is computed by performing a point addition between $3\mathcal{K}$ and $2\mathcal{K}$. Therefore, we only need one point double and seven point addition operations in hardware to compute all the required values. Overall, we significantly reduce the number of point double and addition operations with the use of width-$w$ NAF, and offline and optimized computation for $\mathcal{G}$ and $\mathcal{K}$.

\begin{figure*}[t!]
  \begin{center}
  \vspace{-0.15in}
    \includegraphics[width=6.5in]{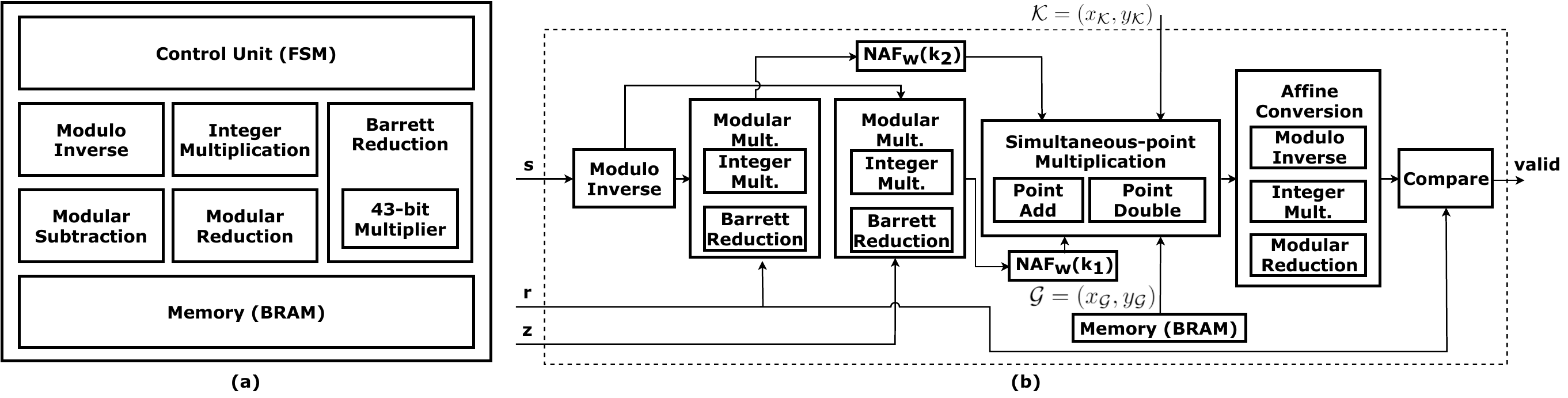}
  \end{center}
  \vspace{-0.15in}
  \caption{Generic ECDSA verification engine: (a) Architecture (b) Data flow.}
  \vspace{-0.15in}
  \label{fig:DFECDSASPM}
\end{figure*}

\subsection{ECDSA Verification Engine}
\label{ECDSAVerfSPM}
Figure~\ref{fig:DFECDSASPM}(a) depicts the architecture of our generic ECDSA verification engine, using the modular and point arithmetic modules described earlier. We instantiate only one module corresponding to a unique modular arithmetic operation to keep the resource utilization low. Figure~\ref{fig:DFECDSASPM}(b) depicts the data flow in the ECDSA verification engine using SPM algorithm. Even with single instantiation of the modular arithmetic modules, we leverage as much parallelism as possible by scheduling different operations in parallel to efficiently utilize the hardware resources. For example, we perform the $NAF_w(k_2)$ conversion in parallel with the second modular multiplication. Similarly, $NAF_w(k_1)$ conversion happens in parallel to the computation on line $1$ in Algorithm~\ref{alg:SPM}. Thus, no additional clock cycles are spent in NAF conversions. From amongst all the modules, integer multiplication and modular reduction are the heavily utilized modules.

It is worthwhile to reiterate that an additional point addition operation is not required in this implementation because it is absorbed in the SPM operation. However, after the multiple-point multiplication operation, we need an additional operation to convert $x$ from projective Chudnovsky coordinate back to affine coordinate for final comparison.  

\section{ECDSA Verification for HL Fabric}
\label{sec:fabric}
In this section, we propose optimizations in the context of permissioned blockchains specifically HL Fabric. We exploit the fact that some parameters are fixed apriori while other parameters are available in advance, and hence both of these can be preprocessed to speedup ECDSA verification operation. More specifically, we take advantage of the fact that the generator point $\mathcal{G}$ is fixed and the public key coordinates $\mathcal{K} = (x_{\mathcal{K}},y_{\mathcal{K}})$ are known well in advance before the ECDSA verification starts. The ProtocolProcessor in Blockchain Machine (see Figure~\ref{fig:BCM}) processes the incoming data and extracts the public key and ECDSA signature information. Therefore, as soon as the public key coordinates are available, we can start processing them. With this goal in mind, we leverage the fixed-point multiplication (FPM) algorithm (refer Algorithm~\ref{alg:FPM}) to perform point arithmetic instead of simultaneous-point multiplication algorithm. 



\begin{algorithm}[!b]
 \caption{Fixed-base NAF windowing method for FPM}
 \begin{algorithmic}[1]
 \renewcommand{\algorithmicrequire}{\textbf{Input:}}
 \renewcommand{\algorithmicensure}{\textbf{Output:}}
 \REQUIRE Window width $w$, $d=256/w$, $k$, point $\mathcal{P}$
 \ENSURE $k\mathcal{P}$  
  \STATE Precompute: $\mathcal{P}_i = 2^{wi}\mathcal{P}$, $0 \le i \le d$
  \STATE Compute NAF($k$)
  \STATE $I = (2^{w+1} - 2)/3$
  \STATE $\mathcal{A} = \infty$, $\mathcal{B} = \infty$
  \FOR{$j$ from $I$ down to $1$}
    \STATE For each $i$ for which $k_i = j$ do $\mathcal{B} = \mathcal{B} + \mathcal{P}_i$
    \STATE For each $i$ for which $k_i = -j$ do $\mathcal{B} = \mathcal{B} - \mathcal{P}_i$
    \STATE $\mathcal{A} = \mathcal{A} + \mathcal{B}$
  \ENDFOR
  \STATE Return$(\mathcal{A})$
 \end{algorithmic}
 \label{alg:FPM}
 \end{algorithm}

The algorithm starts by precomputing various powers-of-$2$ point multiplications for a point $\mathcal{P}$ which is known apriori (for example, when $w=4$, then precomputations will be $\mathcal{P},16\mathcal{P},256\mathcal{P},$ and so on). As the generator point is fixed, we precompute these values offline and store the values in the BRAM on FPGA. For the public key coordinates, we design a separate precompute block, outside of the ECDSA verification engine, which runs binary scalar-point multiplication algorithm (algorithm-$3.27$ in \cite{HG2006}) with point double operations only to precompute the values mentioned earlier (i.e., $\mathcal{P}$,$16\mathcal{P}$, etc). Point addition operations are not required as we are computing power-of-$2$ point multiplications only, which can be computed using successive point double operations. We further optimize the precompute block by reducing the number of point double operations that it needs to perform. For example, if we want to compute $256\mathcal{P}$, we need not start point double operations all the way from $\mathcal{P}$. Instead, we can use the value of $16\mathcal{P}$ that was computed in the previous step, thus reducing the number of point double operations from $wi$ to $w$ in each step $i$ where $0 \le i \le d$. This optimization helps reduce the number of point double operations from over $8000$ to only $252$ when $w=4$ and $d=64$.   

\begin{figure}[b!]
  \begin{center}
    \includegraphics[width=0.45\columnwidth]{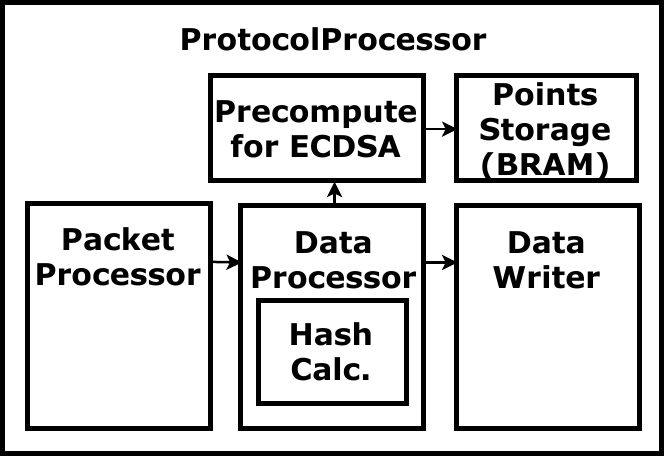}
  \end{center}
  \vspace{-0.15in}
  \caption{Integration of Precompute with ProtocolProcessor.}
  \vspace{-0.10in}
  \label{fig:ProtP}
\end{figure}

\begin{figure*}[t!]
  \begin{center}
    \includegraphics[width=7.0in]{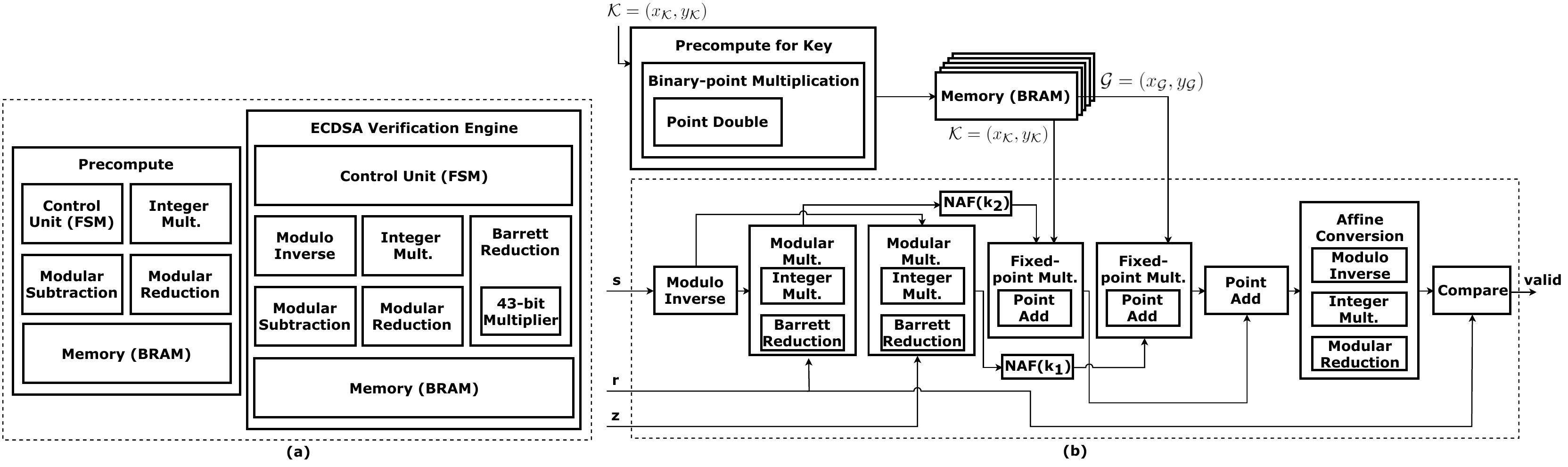}
  \end{center}
  \vspace{-0.15in}
  \caption{ECDSA verification engine with Precompute block: (a) Architecture  (b) Data flow.}
  \vspace{-0.10in}
  \label{fig:DFECDSAFPM}
  \vspace{-0.05in}
\end{figure*}

The precompute block computes the required point values for public key and stores them in the BRAM. The computation within the actual ECDSA verification reduces to lines $4$-$8$ in Algorithm~\ref{alg:FPM}. Consequently, ECDSA verification comprises of just point addition operations with the point double operations moved to precomputation. We use the fixed-base (with base $w = 4$) NAF windowing method to reduce the number of point addition operations. Note that the fixed-point multiplication algorithm is executed twice; once for the generator point $\mathcal{G}$ accessing offline computed point values in line $1$, and second time for the public key $\mathcal{K}$ accessing precomputed point values in line $1$.


Figures~\ref{fig:ProtP} and~\ref{fig:DFECDSAFPM}(a) show the architecture of our HL Fabric-specific ECDSA verification engine. The precompute block is placed inside the ProtocolProcessor to store all the required point values in BRAM. With projective Chudnovsky coordinates, we need about $20KB$ of memory to store all the precomputed point values (including point $\mathcal{G}$ and $\mathcal{K}$). In a permissioned blockchain like HL Fabric, the number of nodes are limited and hence the number of unique identities (public keys) is limited and those identities are known apriori. In a typical HL Fabric network, there may only be tens of unique identities, thus storage of precomputed points will not incur a high memory overhead. The precompute block has its own modular arithmetic modules and the finite state machine (FSM) controlling the movement of data between these modules. The ECDSA verification engine has its own set of modular arithmetic modules. We again instantiated only one module per modular arithmetic operation to keep the resource utilization low. This is beneficial especially in Blockchain Machine where it is desirable to have many ECDSA verification engines within the BlockProcessor module.

Figure~\ref{fig:DFECDSAFPM}(b) depicts the data flow in our ECDSA verification engine with the precompute block. The precomputed points are read from the BRAM by the ECDSA verification engine. The read from BRAM is not a bottleneck as read will happen once in a while and then hundreds of clock cycles are spent on processing the data. Since the number of unique public keys is limited, the precompute block is executed only when a new public key is encountered and the precomputed values are stored in the BRAM. Hence, the precompute block does not become the bottleneck. Both in precompute and ECDSA verification engine, with single instantiation of the modular arithmetic modules, most operations are performed serially. However, we leverage as much parallelism as possible by scheduling different modular operations in parallel. For example, the NAF conversion of $k_2$ happens in parallel with the second modular multiplication involving $k_1$. Similarly, the NAF conversion of $k_1$ also happens in parallel with the first fixed-point multiplication. After the fixed-point multiplication, a point addition operation is required (line $7$ of Algorithm~\ref{alg:ECDSA}) and conversion back to affine coordinates for comparison.
\section{Evaluation}
\label{sec:evaluation}
We designed our ECDSA verification engine and all of its modules in Verilog $2001$ and synthesized the design using Xilinx Vivado design suite 2019.2. For functional verification, we generated the test cases using open-source code from OpenSSL library~\cite{OpenSSL} and also from the actual data, i.e., public key, signature, and hash used in Hyperledger Fabric. We also successfully verified the test vectors~\cite{NIST-DSA-VAL} from NIST for $P$-$256$ (along with SHA-$256$) ECDSA signature verification. We implement the design on Xilinx Alveo U250 FPGA. We synthesize our design on this FPGA board because blockchains are typically deployed on a cloud server with FPGA accelerator card. 
Since our goal is to integrate the ECDSA engine into a blockchain hardware accelerator~\cite{javaid2021blockchain}, which is quite complex and operates at $250$~MHz frequency, we limit the operating frequency of our ECDSA verification engine to $250$~MHz even when it is possible to obtain higher frequencies with DSP blocks like in ~\cite{guneysu2008ultra}.
This restriction on frequency will also enable better scalability within the blockchain accelerator by instantiating multiple ECDSA engines for distributed computation.

Throughout this section, we report frequency (freq.) in MHz, latency in clock cycles, and throughput (TP) in operations per second.
When comparing our work with existing state-of-the-art implementations, we cautiously compare the clock cycles of different designs instead of absolute runtimes to overcome the inherent improvements from upgraded technologies and operating frequencies, and hence provide a fair comparison. In an ideal situation, the existing works should have been implemented on the Alveo U250 FPGA board as well. However, those designs are not open-source and implementation of each requires significant effort. 
Furthermore, we do not present direct comparison results with CPU/GPU implementations because our goal is not to compete with CPU/GPU implementations but to provide the best FPGA implementation that can be used in accelerators for permissioned blockchains (since they are naturally suitable for FPGA based acceleration~\cite{javaid2021blockchain}).

\subsection{Modular Arithmetic}
We start by discussing the area footprint and latency of the individual modular arithmetic modules as listed in Table~\ref{table:HCL}. With all the proposed optimizations, our modular arithmetic modules incur low resource utilization. The integer multiplication module consumes the most LUTs, which can be reduced using BRAM-based optimizations. We, however, leave this optimization for future work. Most modules perform fast computations except for Barrett reduction which has the highest latency because of two serial $258$-bit multiplications. We made this design choice to keep the hardware resource utilization low. Moreover, modular reduction using Barrett reduction is performed only in lines $5$, $6$, and $8$ of Algorithm~\ref{alg:ECDSA}, which is not the critical path (line $7$ is the critical path in ECDSA verification). Note that our modular subtraction module is $1.8 \times$~faster and our modular multiplication (integer multiplication+$P$-$256$ Modular Reduction) is $1.2 \times$~faster than the implementations in state-of-the-art work~\cite{guneysu2008ultra}.

\begin{table}[!t]
\caption{Hardware results of various modular arithmetic modules.}
\vspace{-0.15in}
\begin{center}
  \begin{tabular}{ | l | c | c | c | c |} 
    \hline
    Operation & LUT & FFs & DSP & Latency \\    
    \hline
    \hline
    Modular subtraction & $616$ & $781$ & $1$ & $10$ \\
    \hline
    Integer multiplication & $5471$ & $7980$ & $128$ & $39$ \\
    \hline
    $P$-$256$ Modular Reduction & $2225$ & $789$ & $0$ & $19$ \\
    \hline
    Barrett Reduction & $2130$ & $3597$ & $9$ & $1,552$ \\
    \hline
    Modulo Inverse & $3503$ & $1313$ & $0$ & $~550$ \\
    \hline
\end{tabular}
\label{table:HCL}
\vspace{-0.10in}
\end{center}
\end{table}

\vspace{-0.10in}
\begin{table}[!t]
\caption{Performance comparison of point arithmetic.}
\vspace{-0.15in}
\begin{center}
  \begin{tabular}{ | l | c| c| c | c|} 
    \hline
    Operation & Platform & Freq. & Latency & TP \\    
    \hline
    \hline
    PA [Our work] & Alveo U250 & $250$ & $622$ & $402$K \\
    \hline
    PA~\cite{guneysu2008ultra} & Virtex-4 & $375$ & $980$ & $382$K  \\
    \hline
    \hline
    PD [Our work] & Alveo U250 & $250$ & $435$ & $574$K \\
    \hline
    PD~\cite{guneysu2008ultra} & Virtex-4 & $375$ & $700$ & $535$K \\
    \hline
    \hline
    PM [Our work] & Alveo U250 & $250$ & $190,976$ & $1,309$ \\
    \hline
    PM~\cite{guneysu2008ultra} & Virtex-4 & $375$ & $303,450$ & $1,236$ \\
    \hline
    PM~\cite{vliegen2010compact} & Virtex-2 Pro & $108.2$ & $451,733$ & $240$ \\
    \hline
    PM~\cite{kudithi2019high} & Virtex-7 & $124.2$ & $462,520$ & $268$ \\
    \hline
    PM~\cite{mentens2007secure} & Virtex-2 Pro & $67$ & $567,500$ & $118$ \\
    \hline
    PM~\cite{mcivor2004fpga} & Virtex-2 & $39.5$ & $960,000$ & $41$ \\
    \hline
    PM~\cite{schinianakis2008rns} & Virtex-E & $39.7$ & $987,500$ & $40$ \\
    \hline
    \hline
    SPM [Our work] & Alveo U250 & $250$ & $231,406$ & $1,080$ \\
    \hline
    SPM~\cite{guneysu2008ultra} & Virtex-4 & $375$ & $366,905$ & $1,022$ \\
    \hline
    SPM-NAF [Our work] & Alveo U250 & $250$ & $181,024$ & $1,381$ \\
    \hline
\end{tabular}
\label{table:PAML}
\vspace{-0.15in}
\end{center}
\end{table}

\subsection{Point Arithmetic}
We first evaluate the performance of our point addition (PA) and point double (PD) operations. For a fair comparison, we also report the latencies from state-of-the-art FPGA-based work~\cite{guneysu2008ultra}, which also uses projective Chudnovsky coordinates like our implementation. We observe that both of our point add and double operations are \textasciitilde$1.6 \times$ than their point add and double operations in terms of clock cycles (refer Table~\ref{table:PAML}). Note that their design involves a dual clock which is much more complicated to implement than our design, which uses only a single clock throughout the entire design. 
Moreover, the authors in \cite{guneysu2008ultra} focus on optimizing only the point arithmetic, and do not implement the entire ECDSA verification algorithm. This makes it much easier for their stand-alone point arithmetic modules to run at higher frequencies.  

Now we compare the latencies of our scalar point multiplication (PM) with state-of-the-art works that implement scalar point multiplication (refer Table~\ref{table:PAML}). We observe that our PM is \textasciitilde$1.6 \times$ to \textasciitilde$5 \times$~faster than these existing works in terms of clock cycles. As most of these prior implementation were done using binary double and add algorithm, our PM is also implemented using the same approach. Note that implementation done by Kudithi et al.~\cite{kudithi2019high} works with affine coordinates while rest of the works use projective coordinates. Next, we compare the latencies of our simultaneous-point multiplication (SPM) operation. As mentioned in Section~\ref{PA}, we perform SPM operation using width-$4$ NAF approach. However, for fair comparison with \cite{guneysu2008ultra}, which uses a binary representation, we estimated the latencies using the number of operations performed in Algorithm~\ref{alg:SPM} with binary representation. More importantly, our SPM with width-$4$ NAF incurs about half the latency owing to our faster modular arithmetic modules.

\subsection{ECDSA Verification Engine}
Table~\ref{table:ECDSA} presents the hardware resource utilization and latency of the ECDSA verification engine. Our ECDSA verification engine takes \textasciitilde$190,000$~clock cycles for a single signature verification leading to a throughput of $1,315$~signature verifications per second. Although many prior works have accelerated point arithmetic on FPGA, most works did not implement the entire ECDSA verification algorithm. We found only two relevant comparable works in literature that accelerated ECDSA signature verification for NIST $P$-$256$ on FPGA, which are reported in Table~\ref{table:ECDSALT}. Glas et al.~\cite{glas2011prime} reported the implementation results of the complete signature generation and verification unit on a Xilinx XC5VLX110T Virtex-5 FPGA. Their signature verification unit includes a hash generator ip which incurs a latency of $68$~clock cycles that we have adjusted accordingly for a fair comparison. Their design achieves a throughput of $110$~signature verifications per second that is about $12 \times$ lower than the throughput of our ECDSA verification engine.   

\begin{table}[!b]
\caption{Hardware results of ECDSA engine.}
\vspace{-0.15in}
\begin{center}
  \begin{tabular}{ | l | c | c | c | c | c |} 
    \hline
    Design & LUT & FFs & DSP & BRAM & Latency \\    
    \hline
    \hline
    ECDSA verf. & $24394$ & $10961$ & $137$ & $5$ & $190,000$ \\
    \hline
\end{tabular}
\label{table:ECDSA}
\vspace{-0.15in}
\end{center}
\end{table}

\begin{table}[!b]
\caption{Performance comparison of ECDSA engine.}
\vspace{-0.15in}
\begin{center}
  \begin{tabular}{ | l | c | c | c | c |} 
    \hline
    Work & Platform & Freq. & Latency & TP \\    
    \hline
    \hline
    Our work & Alveo U250 & $250$ & $190,000$ & $1,315$ \\
    \hline
    \cite{glas2011prime} & Virtex-5 & $50$ & $454,140$ & $110$ \\
    \hline
    \cite{knevzevic2016low} & Virtex-4 & $33.3$ & $69,972$ & $475$ \\
    \hline
\end{tabular}
\label{table:ECDSALT}
\vspace{-0.10in}
\end{center}
\end{table}

Kne{\v{z}}evi{\'c} et al.~\cite{knevzevic2016low} did an ASIC implementation for ECDSA verification, but reported performance numbers by synthesizing their design on a Xilinx Virtex-4 FPGA. Their design results in a throughput of $475$~signature verifications per second, which is \textasciitilde$2.8 \times$ lower than the throughput of our ECDSA verification engine. 
Note that this design is primarily an ASIC implementation, which means that it has a different optimization flow, resulting in a much lower operating frequency and requiring less clock cycles. Therefore, we do not compare frequency and latency for this design, but compare the throughput for a fair comparison.
We do not compare the hardware resource utilization as prior designs use different FPGA boards for implementation. Moreover, we cannot estimate the hardware cost of these existing designs~\cite{glas2011prime,knevzevic2016low} for Alveo U250 FPGA as these designs are not open-sourced.

\subsection{ECDSA Verification Engine for HL Fabric}
Table~\ref{table:ECDSABM} presents the hardware resource utilization and latency of our ECDSA verification engine with precompute block. We observe that the precompute block incurs a latency of \textasciitilde$120,000$ clock cycles. Then, the actual ECDSA verification engine requires only \textasciitilde$92,000$~clock cycles to perform a single signature verification. With this approach, we achieve a throughput of $2,717$~signature verifications per second. 

We also evaluate the HL Fabric-specific ECDSA verification engine in the context of Blockchain Machine. Table~\ref{table:BMECDSA} presents the throughput (transactions per second) of Blockchain Machine with both the generic and Hyperledger Fabric-specific ECDSA verification engines. We observe a $2 \times$ improvement in throughput with our precompute optimization. We also change the number of ECDSA verification engines from $4-10$, and observe that both types of engines scale the throughput in a similar trend (\textasciitilde1.57$\times$ improvement).

\begin{table}[!t]
\caption{Hardware results of ECDSA engine ($250$~MHz).}
\vspace{-0.15in}
\begin{center}
  \begin{tabular}{ | l | c | c | c | c | c | c |} 
    \hline
    Design & LUT & FFs & DSP & BRAM & Latency & TP  \\    
    \hline
    \hline
    Precompute & $14088$ & $1417$ & $129$ & $15$ & $120,000$ & - \\
    \hline
    ECDSA eng. & $21759$ & $5625$ & $137$ & $15$ & $92,000$ & $2,717$ \\
    \hline
\end{tabular}
\label{table:ECDSABM}
\vspace{-0.15in}
\end{center}
\end{table}

\begin{table}[!t]
\caption{TP of Blockchain Machine with ECDSA engines.}
\vspace{-0.15in}
\begin{center}
  \begin{tabular}{ | l | c | c |} 
    \hline
    No. of Engines & Generic ECDSA & HL Fabric ECDSA \\
    \hline
    \hline
    4 & $1,290$~tps & $2,650$~tps \\
    \hline
    7 & $2,525$~tps & $5,200$~tps \\
    \hline
    10 & $3,650$~tps & $7,520$~tps \\
    \hline
\end{tabular}
\label{table:BMECDSA}
\vspace{-0.15in}
\end{center}
\end{table}
\section{Conclusion}
\label{sec:conclusion}
\vspace{-0.05in}
In this work, we focused on an FPGA-based efficient implementation of ECDSA signature verification, in order to improve the performance of permissioned blockchains that aim to use FPGA-based accelerators. We presented several FPGA-specific algorithmic optimizations for modular arithmetic modules. With efficient utilization of DSP blocks on FPGA, we showed on average $1.5 \times$ speedup compared to \cite{guneysu2008ultra}. We also presented optimizations for SPM and FPM algorithms, and used projective Chudnovsky coordinate with optimal width NAF representations. With these optimizations, we observed on average $3.2 \times$ speedup in our point arithmetic operations compared to~\cite{guneysu2008ultra,vliegen2010compact,kudithi2019high,mentens2007secure,mcivor2004fpga,schinianakis2008rns}. Our ECDSA verification engine, using these modular and point arithmetic modules, performs a signature verification in $760\mu s$ resulting in a throughput of $1,315$ verifications per second, which is \textasciitilde$2.5\times$ faster than~\cite{glas2011prime,knevzevic2016low}. With HL Fabric-specific optimizations, our ECDSA verification engine can perform a signature verification in $368\mu s$ resulting in a throughput of $2,717$ verifications per second.

\bibliographystyle{ACM-Reference-Format}
\bibliography{References/References}

\end{document}